\documentclass[a4paper,10pt,twoside]{cpc-hepnp}
\usepackage{CJK,upgreek,fancyhdr}
\usepackage{multicol}
\usepackage{graphicx}
\usepackage{booktabs}
\usepackage{amssymb,bm,mathrsfs,bbm,amscd}
\usepackage[tbtags]{amsmath}
\usepackage{lastpage}
\usepackage{color}

\begin{document}
\begin{CJK*}{GB}{gbsn}

\fancyhead[R]{\small To be published in \emph{Chinese Physics C}}
\fancyfoot[C]{\small 010201-\thepage}

\footnotetext[0]{Received 10 June 2016}

\title{Design of FELiChEM, the first infrared free-electron laser user facility in China\thanks{Supported by National Natural Science
Foundation of China (21327901) }}

\author{%
      He-Ting Li(ÀîºÍÍ¢)$^{1)}$\email{liheting@ustc.edu.cn}%
\quad Qi-Ka Jia(¼ÖÆô¿¨)
\quad Shan-Cai Zhang(ÕÅÉƲÅ)$^{2)}$\email{shancai@ustc.edu.cn}%
\quad Lin Wang(ÍõÁÕ)
\quad Yong-Liang Yang(ÑîÓÀÁ¼)
}
\maketitle

\address{%
National Synchrotron Radiation Laboratory, University of Science and Technology of China, Hefei, 230029, Anhui, China\\
}

\begin{abstract}
FELiChEM is a new experimental facility under construction at University of Science and Technology of China (USTC), whose core device is two free electron laser oscillators generating middle-infrared and far-infrared laser and covering the spectral range of 2.5-200 $\rm{\mu}$m. It will be a dedicated infrared light source aiming at energy chemistry research. We present the brief design of FEL oscillators with the emphasis put on the middle-infrared oscillator. Most of the basic parameters are determined and the anticipated performance of the output radiation is given. The first light of FELiChEM is targeted for the end of 2017.
\end{abstract}



\footnotetext[0]{\hspace*{-3mm}\raisebox{0.3ex}{$\scriptstyle\copyright$}2013
Chinese Physical Society and the Institute of High Energy Physics
of the Chinese Academy of Sciences and the Institute
of Modern Physics of the Chinese Academy of Sciences and IOP Publishing Ltd}%

\begin{multicols}{2}

\section{Introduction}

Lasers are ubiquitous sources of coherent electromagnetic radiation over a wide portion of the spectrum, from the near-infrared (around 10 ${\rm{\mu m}}$) down to the ultraviolet (around 200 nm) \cite{lab1}. They are compact, inexpensive, and easily available. However, in the part of the electromagnetic
spectrum from around 10 ${\rm{\mu m}}$ to 1 mm (typically called the near-infrared to the terahertz region), conventional lasers are not easily available, which have the disadvantages of limited wavelength tunability and/or low intensity, such as the gas laser, the diode laser and so on. Therefore, there is considerable interest in alternate sources of coherent radiation in this electromagnetic spectrum.

A free electron laser (FEL) is a device that transforms the kinetic energy of a relativistic electron beam into electromagnetic (EM) radiation when the electron beam goes through a periodically alternating magnetic field. It can provide coherent radiation in any part of the electromagnetic spectrum.
Even more, the wavelength can be continuously tuned, the intensity can be very high, and the pulse length can be very short. These attributes make the FEL extremely attractive as a coherent radiation source.
Nowadays infrared FELs are used worldwide as user facilities, such as the CLIO in France \cite{lab2,lab3}, the FHI-FEL in German \cite{lab4,lab5}, the FELIX in Netherlands \cite{lab6,lab7} and so on. The FELs are especially efficient in infrared since they produce high power laser pulses, typically MW level in peak power and up to hundreds of mJ level in macropulse energy, and are tunable across a large wavelength range, typically 1 to 2 decades.

Following the Dalian Coherent Light Source (DCLS) \cite{lab8}, in 2014, the project of infrared laser for fundamental of energy chemistry, named FELiChEM, was approved under the financial support of Natural Science Foundation of China, and beginning to build up an infrared FEL (IR-FEL) in Hefei. The National Synchrotron Radiation Laboratory (NSRL) of USTC is responsible for the design, construction and commissioning of IR-FEL apparatus. It will be a dedicated experimental facility aiming at energy chemistry research, whose core device is a free electron laser (FEL) generating 2.5-200 $\rm{\mu}$m laser for photo excitation, photo dissociation and photo detection experimental stations. Similar as the FHI-FEL, two oscillators driven by one RF Linac will be used to generate mid-infrared (MIR) (2.5-50 $\rm{\mu}$m) and far-infrared (FIR) (40-200 $\rm{\mu}$m) lasers.
 FELiChEM has clear scientific targets which bring many special requirements on the FEL facility. For the excitation and dissociation of molecules and clusters, the users demand the peak and average power of IR-FEL as high as possible, and some others require a special wavelength range that can be tuned continuously (just tuning the undulator gap). All these demand should be considered and satisfied in the design.

We once considered using the single-pass FEL amplifier driven by prebunched electron bunch train as described in Ref. \cite{lab9,lab10,lab11}. However, it is difficult to reach the demanded radiation pulse energy and it is not convenient to tune the FEL wavelength in a broad range. Furthermore, this single-pass FEL requires a higher electron beam quality. Finally, the traditional FEL oscillator is selected, which has been theoretically and experimentally studied in China for decades \cite{lab12,lab13,lab14,lab15,lab16}.

In this paper, We present the brief design of FEL oscillators that delivers the required performance for this device, including the undulator, the optical resonators, and the electron beam. Then we show the anticipated performance of the output radiation. Emphasis is put on the design of the MIR-FEL oscillator. We finally summarize in the last section.

\section{Layout of FELiChEM}
As shown in Fig. 1, the IR-FEL is composed of two FEL oscillators driven by one electron Linac. Two accelerating tubes (A1, A2) are used to accelerate the electron beam to the maximum energy of 60 MeV. Between the first and the second accelerating tube, a four-dipole magnetic chicane is designed as an optional operation condition. Its purpose is to reduce the micro-pulse length and increase the peak current of the electron bunch for the short-wavelength FEL, and for the long-wavelength FEL, it also can increase the micro-pulse length to suppress the slippage effect. Since the first accelerating tube is capable of reaching the required electron energy for FIR-FEL, we extract the beam into the FIR oscillator after the first accelerating tube, for leaving enough space between the two oscillators. Between the electron Linac and FEL oscillators, the achromatic transfer lines are designed, where energy collimators will be used to eliminate the electrons with large energy spread, and the quadrupole doublets will be used to adjust transverse matching between the electron beam and the laser beam inside the oscillators.

The IR-FEL facility will be placed in a 16m$\times$10m$\times$3.2m hall with 2.8-meters-thick shield walls.
\end{multicols}
\begin{center}
\includegraphics[width=12cm]{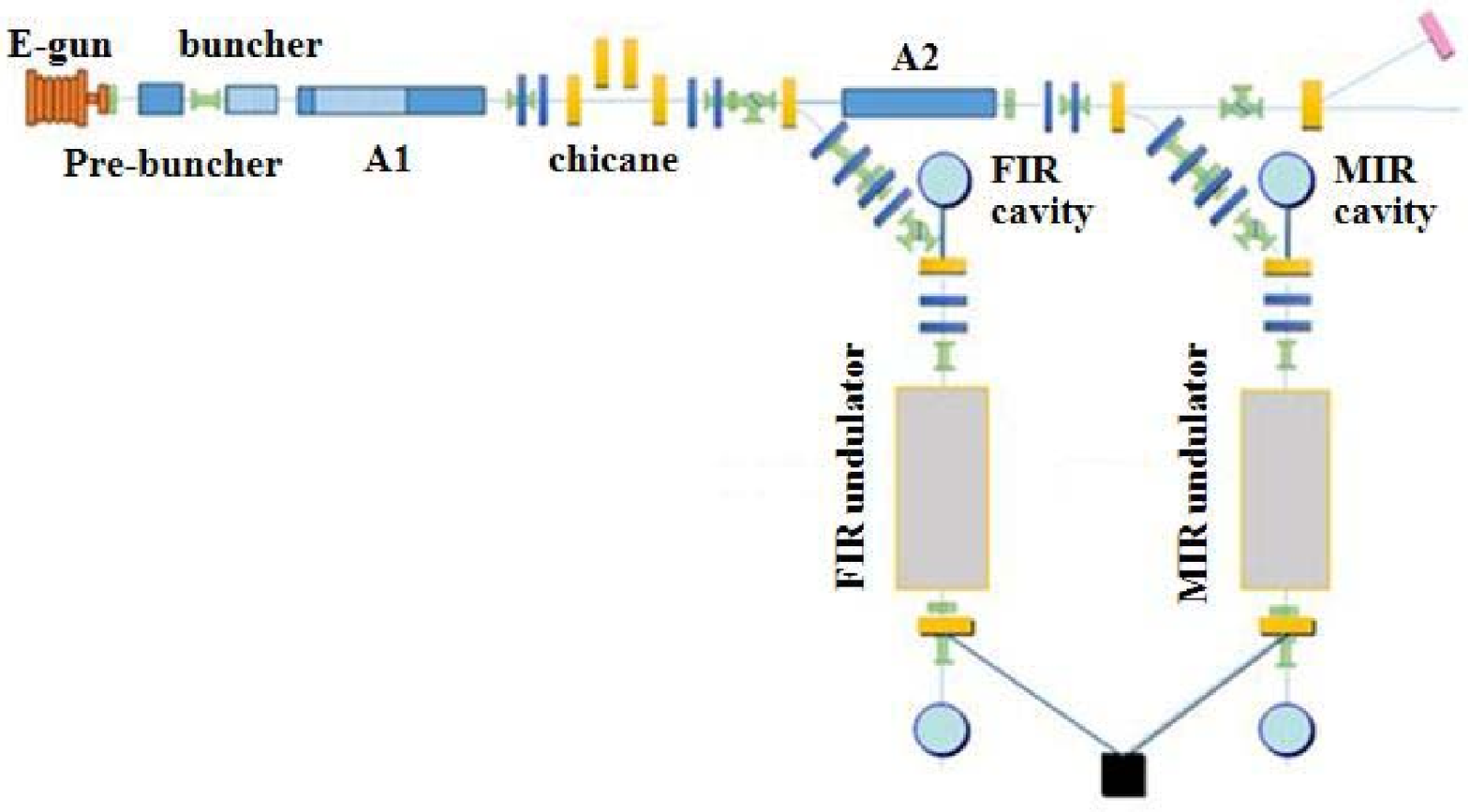}
\figcaption{\label{fig1}   Layout of FELiChEM. }
\end{center}
\begin{multicols}{2}

\section{Design methodology}

Theoretical formula and numerical simulations are used in the parameter optimization of FELiChEM, where parameter dependences are more evident in theoretical formulas while numerical simulation is more accurate and more reliable.

 For an FEL oscillator, the intra-cavity power $P_m $ after $m$ passes can be given as
  \begin{eqnarray}
P_m  = P_0 \prod\limits_{i = 1}^m {e^{g_i  - \alpha } }
\end{eqnarray}
where $P_0 $ is the power of the undulator spontaneous radiation, $g_i$ is the gain factor of the $i$th pass, and $\alpha$ is the total loss including outcoupling, diffraction loss, and loss on the mirrors. When the gain is equal to the total loss, the radiation becomes saturated. Evidently, the gain factor and the total loss are the key factors determining the FEL performance. In the design of an FEL oscillator, one should make the best of a high gain and a low loss. The small-signal gain is,
  \begin{eqnarray}
g_{ss}  =  - (4\pi N_u \rho )^3 \frac{d}{{d\tau }}{\rm{sinc}}^{\rm{2}} (\tau /2)
\end{eqnarray}
where $\rho$ is the FEL pierce parameter, $N_u$ is the period number of the undulator, and $\tau  = 4\pi N_u \eta _0$, $\eta _0$ is the deviation of the electron energy. For a real beam, the beam energy spread, beam emittance and limited electron bunch length will affect the FEL gain, which can be described by adding several empirical modifying factors as [17, 18]
  \begin{eqnarray}
G = f_{\sigma _\gamma  } f_{\varepsilon _y } f_{\mu _c } g_{ss}
\end{eqnarray}

Nowadays, the numerical computation is capable of simulating the evolution of the optical field in an FEL oscillator. In our work, for accuracy and reliablility, three-dimension code OPC (Optical Propagation Code) [19], which can simulate the propagation of the optical field through an optical cavity and cooperates with Genesis code [14] to simulate three-dimension FEL gain process, is used for detailed calculation. This numerical method has been checked by comparing the simulations with the observations of the oscillator at the Thomas Jefferson National Accelerator Facility and the simulation results are in substantial agreement with the experiment [20].

\section{MIR-FEL oscillator}
Normally shorter-wavelength FEL has tighter requirements on the electron beam and the system parameters of the undulator and optical resonator. That is why we put emphasis on the design of the MIR-FEL oscillator. This design determines the optimized parameters of the undulator, the optical resonator and the electron beam. Based on these designed parameters, we simulate the MIR oscillator by OPC code and show the FEL performance.

\subsection{Undulator}

In a free electron laser, the electron beam and radiation wave interact continually under the resonant condition. For a uniform undulator with period~$\lambda _u$~and strength parameter~$K$, the fundamental resonant wavelength is
\begin{eqnarray}
\lambda _s  = \frac{{\lambda _u }}{{2\gamma ^2 }}(1 + \frac{{K^2 }}{2})
\end{eqnarray}
where~$\gamma$~is the Lorentz factor of electron. The simplest way to tune the FEL wavelength is varying the undulator parameter~$K$, corresponding to varying the undulator gap. However, the maximum undulator parameter~$K _ {max}$ is limited by the undulator period and the maximum peak field at the minimum gap as $K  = 0.934 B_0 [\rm T] \lambda _u [\rm {cm}]$.

Therefore, we need to determine the undulator period appropriately, so that we can achieve the FEL in the objective wavelength range with appropriate electron energies, and furthermore, combining with the selection of undulator length we can get enough high FEL gains at all the wavelengths. In addition, we have to consider the continuous tunability of the FEL wavelength, especially for the range required by users, e.g. 15-50$\rm{\mu}$m.

The designed undulator parameters for MIR-FEL are given in Table 1. We select a planar hybrid undulator adopting NdFeB permanent magnet with a remanence of~$\emph{B}_r=1.2$~T, and Figure 2 shows the peak magnetic field $B_0$ and undulator parameter~$K$ as functions of undulator gap.
Under this condition, the radiation wavelength tunability with different electron beam energy is shown in Fig.3, from which one can find that the continuous tunability can easily reach 300\%. Note that the usages of a long undulator period and the maximum electron beam energy of 60 MeV are mainly for enhancing the performance of radiation around 2.5 $\rm{\mu}$m.

\begin{center}
\tabcaption{ \label{tab1}  Undulator parameters for MIR-FEL.}
\footnotesize
\begin{tabular*}{80mm}{c@{\extracolsep{\fill}}ccc}
\toprule Parameter  & Specification & Unit \\
\hline
Period\hphantom{00}  & 46\hphantom{0} & mm \\
Period number\hphantom{00} & 50\hphantom{0}& -   \\
Min. gap\hphantom{0} & 16\hphantom{0} & mm  \\
Peak magnetic field & 0.1-0.72 \hphantom{0} & T \\
Strength parameter $K$ & 0.5-3.2\hphantom{0}& -  \\
\bottomrule
\end{tabular*}
\end{center}

\begin{center}
\includegraphics[width=7cm]{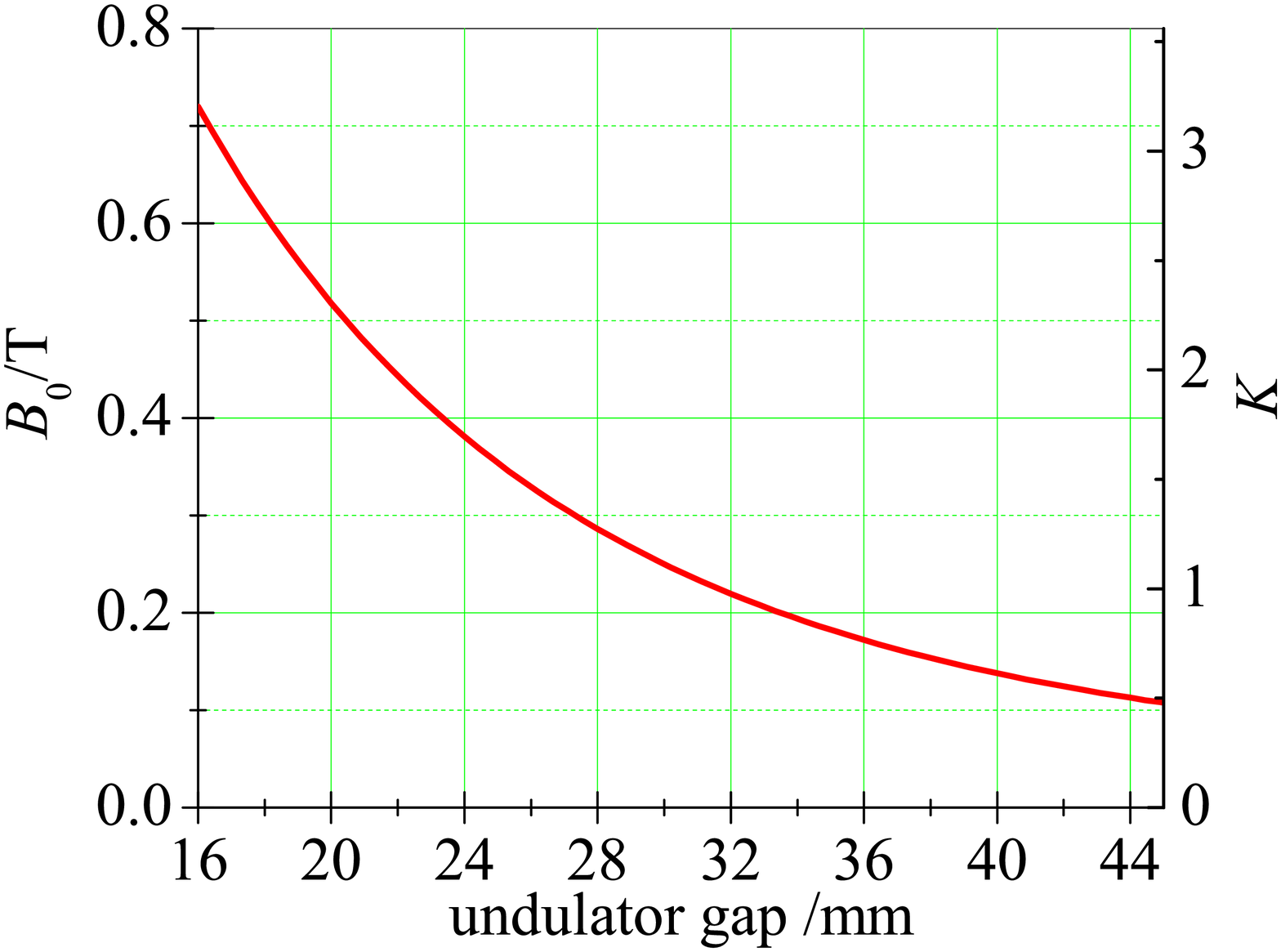}
\figcaption{\label{fig2}   The peak magnetic field $B_0$ and undulator parameter~$K$ as functions of undulator gap. }
\end{center}

\begin{center}
\includegraphics[width=7cm]{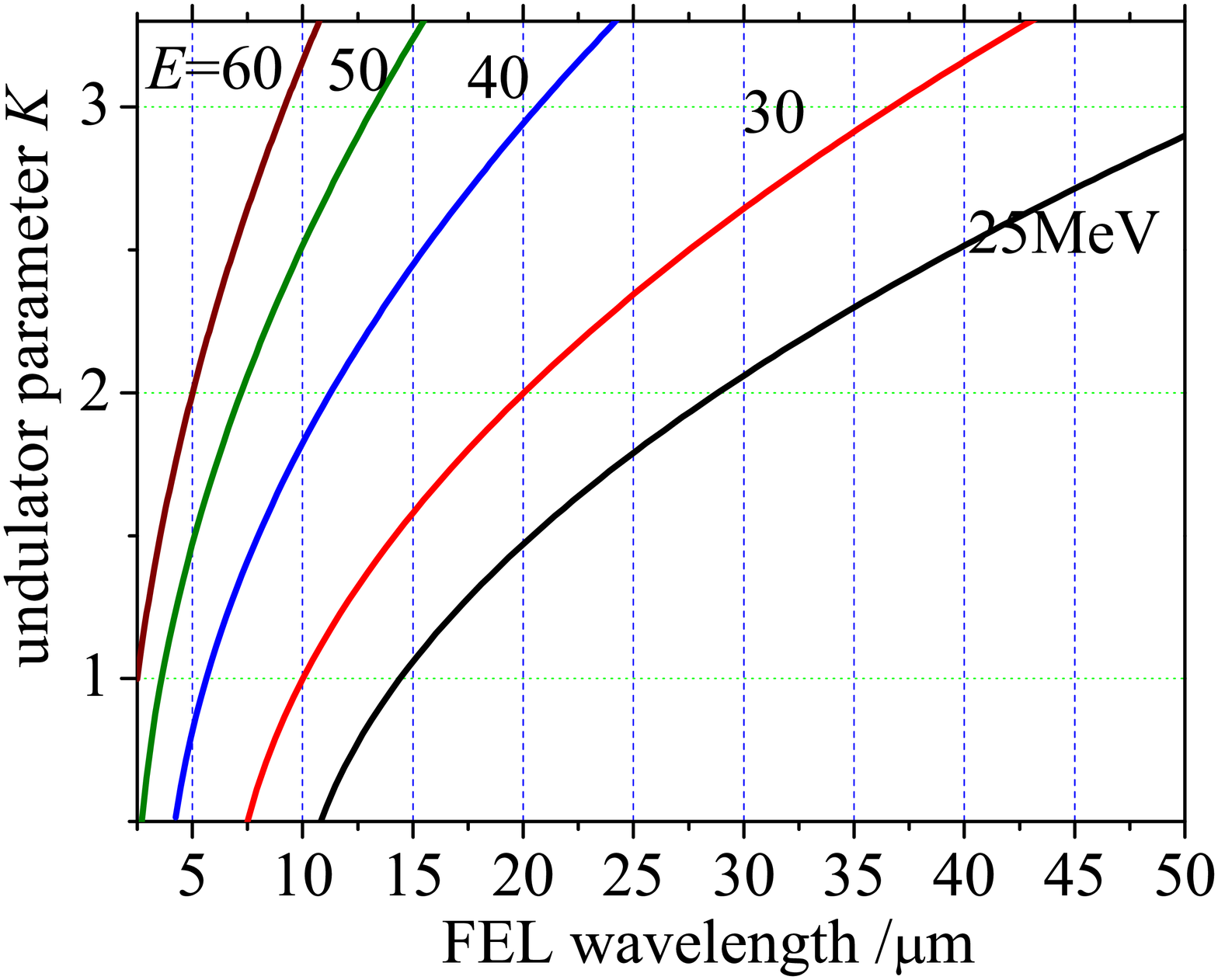}
\figcaption{\label{fig3}   The wavelength tunability with different electron beam energy for MIR-FEL. }
\end{center}

\subsection{Optical resonator}

Optical resonator is a key component for an FEL oscillator, which cumulates the optical field generated by the multiple passes of electron bunches through the undulator, and output a small part of the stored field by using hole outcoupling or some other methods. It consists of two focusing mirrors located on the undulator axis facing each other. There are several key parameters for the optical cavity, such as cavity length, reflectivity of mirrors, curvature radius of mirrors and outcoupling hole size, and so on. The cavity length is determined by following factors, such as the time structure of the electron beam, installation space of other elements, the saturation time of the radiation field, the requirements of optical beam sizes on mirrors, etc. The curvature radius of the mirrors determines the Rayleigh length, stability factor, optical beam size on the mirrors, and the matching of the electron beam and the optical beam. The size of the outcoupling hole contributes to the single-pass loss and then affects the saturation process. When the FEL wavelength varies in a broad range, these relations become more complicated.
\begin{center}
\tabcaption{ \label{tab1}  Parameters for MIR optical resonator.}
\footnotesize
\begin{tabular*}{80mm}{c@{\extracolsep{\fill}}ccc}
\toprule Parameter  & Specification & Unit \\
\hline
Cavity length\hphantom{00}  & 5.04\hphantom{0} & m \\
Rayleigh length\hphantom{0} & 0.77\hphantom{0} & m  \\
Curvature radius of mirrors\hphantom{00} & 2.756\hphantom{0}& m   \\
Reflectivity & 99\% \hphantom{0} & - \\
Diameters of mirrors & 50\hphantom{0}& mm  \\
Diameters of coupling hole & 1.0, 1.5, 2.5, 3.5\hphantom{0}& mm  \\
\bottomrule
\end{tabular*}
\end{center}

In FELiChEM, the resonator is symmetric and the 2.3 m long undulator is located in the center. The radiation is outcoupled from the downstream mirror with a hole. For MIR oscillator, there is no waveguide, because the FEL wavelengths is short so that the cross section of an open resonator mode is small enough to ensure both high laser gain and small diffraction losses. The designed parameters for MIR oscillator are given in Table 2. With the 5.04 meters long cavity, all the micropulse repetition rate can wok well, and the next longer cavity length is 7.56 m. The rayleigh length is designed to be one third of the undulator length (0.77 m).

The optimized rayleigh length for different wavelength is also different, for which we once considered using more than one set of mirrors to provide different Rayleigh lengths. However, to keep the optical beam waist in the middle of the undulator, one has to replace both the upstream and downstream mirrors to obtain a new rayleigh length. This will bring us many technical problems, for instance, the demand of an online re-alignment for the mirrors and the whole optical beam line. Thus, all the mirrors used in the same oscillator are with the same curvature radius. Namely the rayleigh length can not be changed. Under this condition, we just need to scan the attitude of the downstream mirrors to find the optical line after switching mirrors. The selection of the 0.77-meter rayleigh length is mainly for the consideration of 2.5 $\rm{\mu}$m FEL. Since the gain at the wavelength around 2.5 $\rm{\mu}$m is low, a small rayleigh length can make the spot sizes on the mirrors larger to reduce the outcoupling rate. This rayleigh length may induce a larger diffraction loss on the vacuum chamber for the wavelength region of 20-50 $\rm{\mu}$m, however, it can be accepted because the gains at these wavelengths are high enough. Figure 4 shows the intra-cavity optical modes for several typical wavelengths, from which one can note that 2.5 $\rm{\mu}$m FEL has a small spot size on the mirror, and at this moment the outcoupling rate of 1-mm-diameter hole is about 8\%. One also can find that for the wavelength longer than 20 $\rm{\mu}$m there is a little diffraction loss on the undulator chamber. According to the full beam size on the mirrors, the diameter of the mirrors is determined to be 50 mm.

\begin{center}
\includegraphics[width=7cm]{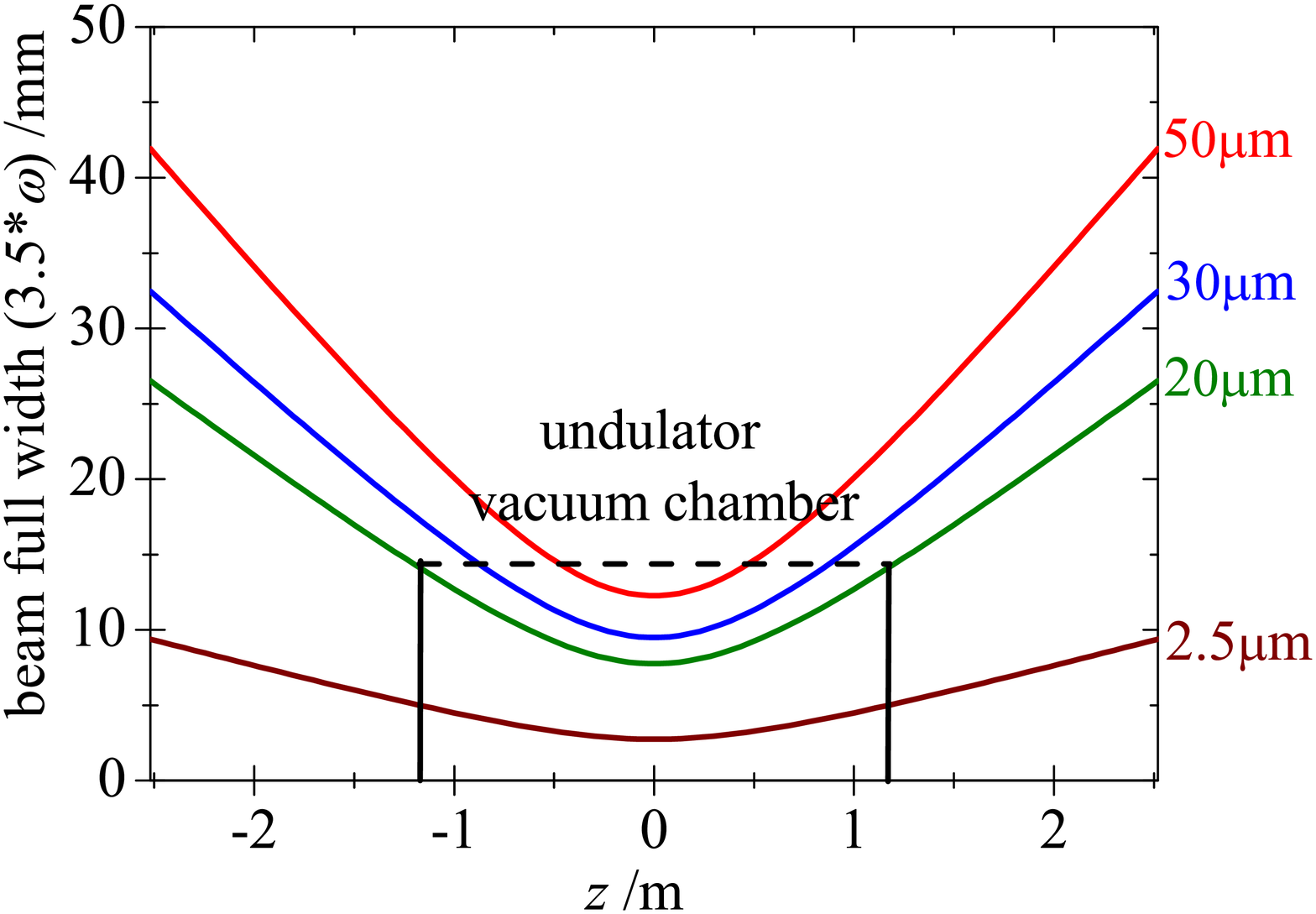}
\figcaption{\label{fig4}   The optical modes in the MIR optical cavity. }
\end{center}

In a hole-coupling oscillator, usually there is a problem that the hollow transverse mode after outcoupling results a significant reduction of output intensity, because of the weak interaction between the hollow mode and the electron beam. That is why FHI-FEL adopts an asymmetrically cavity with the undulator position being offset by 0.5 m from the cavity center away from the outcoupling mirror \cite{lab21}. However, in FELiChEM, benefiting from the downstream outcoupling, the hollow mode travels through the cavity, after reflected by the upstream mirror, then get to the entrance of the undulator. In this long distance propagation, the hollow mode recovers to gaussian mode. Figure 5 gives the 40 $\rm{\mu}$m FEL with the largest hole of 3.5 mm diameter as an example.

Another consequence of hole-coupling is that different hole diameters are needed to optimize the FEL performance at different wavelengths. Therefore, four mirrors with hole diameters of 1.0, 1.5, 2.5, 3.5 mm will be installed in the downstream cavity chamber and can be switched rapidly. The precise cavity length detuning will be realized by adjusting the longitudinal position of the upstream mirror.

\begin{center}
\includegraphics[width=8cm]{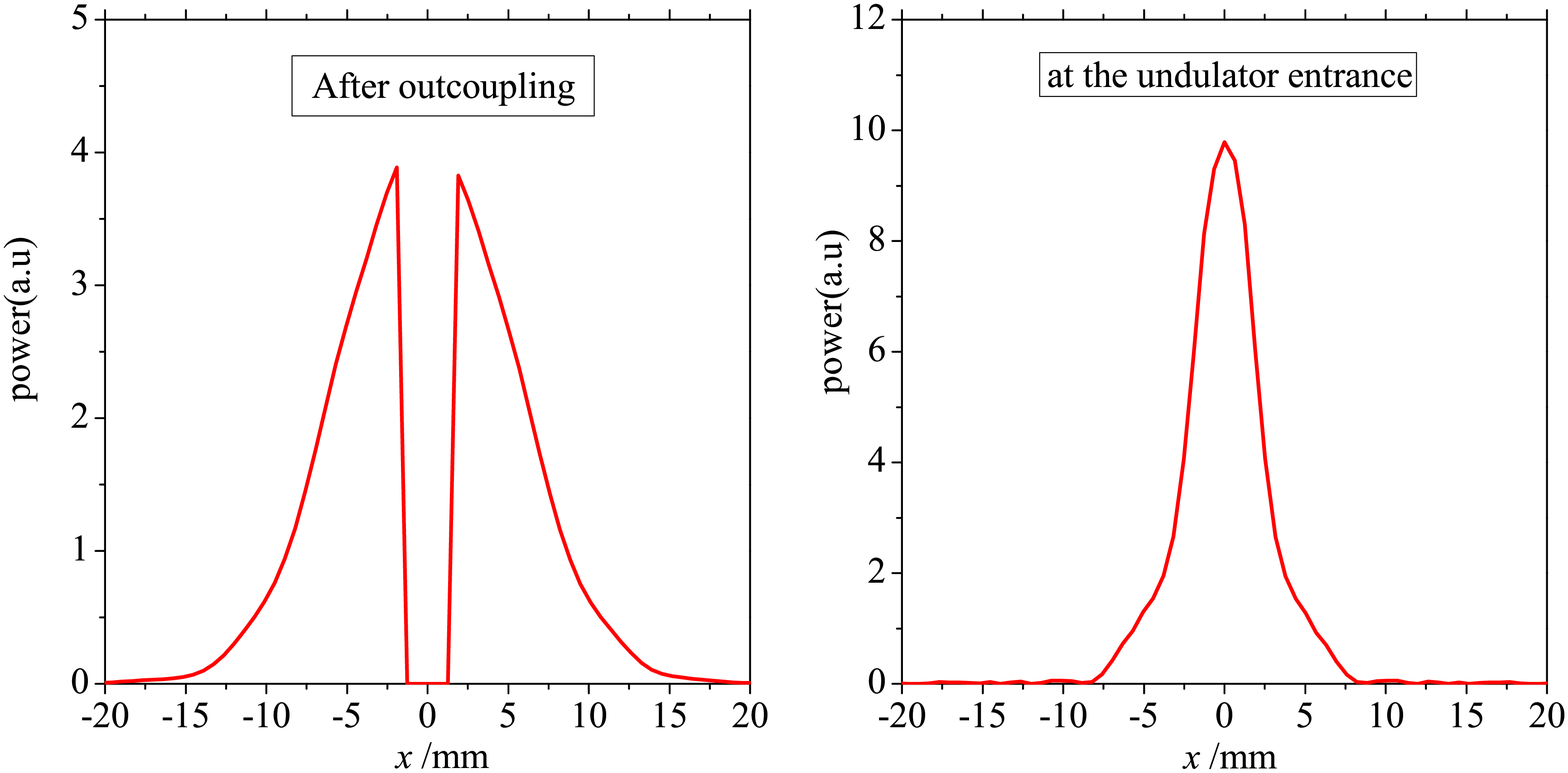}
\figcaption{\label{fig5}   The transverse optical modes of 40 $\rm{\mu}$m FEL with a outcoupling hole of 3.5 mm diameter. {\bf Left}: on the downstream mirror after outcoupling; {\bf Right}: at the undulator entrance.}
\end{center}

\subsection{Electron beam}
The electron source determines the roughly possible quality of the electron beam. We select the thermionic electron gun as the electron source. Using special gate control system, the electron bunch chains will be extracted from the gun, with optional micro-pulse repetition frequency of 476/238/119/59.9 MHz, and the charge of larger than 1 nC. This high bunch charge benefits the single-pass gain and make a lower requirements on the energy spread and emittance.

After optimization, the parameters of the electron beam for MIR-FEL are given in Table 3. On one hand, these parameters satisfy the FEL requirements, and on the other hand, they can be achieved by the RF Linac based on the Linac design.

\begin{center}
\tabcaption{ \label{tab1}  Parameters of electron beam for MIR-FEL.}
\footnotesize
\begin{tabular*}{80mm}{c@{\extracolsep{\fill}}ccc}
\toprule Parameter  & Specification & Unit \\
\hline
Energy\hphantom{00}  & 25-60\hphantom{0} & MeV \\
Energy spread\hphantom{00} & $<$240\hphantom{0}& keV   \\
Emittance\hphantom{0} & $<$30\hphantom{0} & mm$\cdot$mrad  \\
Bunch charge & 1.0 \hphantom{0} & nC \\
Bunch length (rms) & 2-5\hphantom{0}& ps  \\
\bottomrule
\end{tabular*}
\end{center}

\end{multicols}
\begin{center}
\includegraphics[width=7cm]{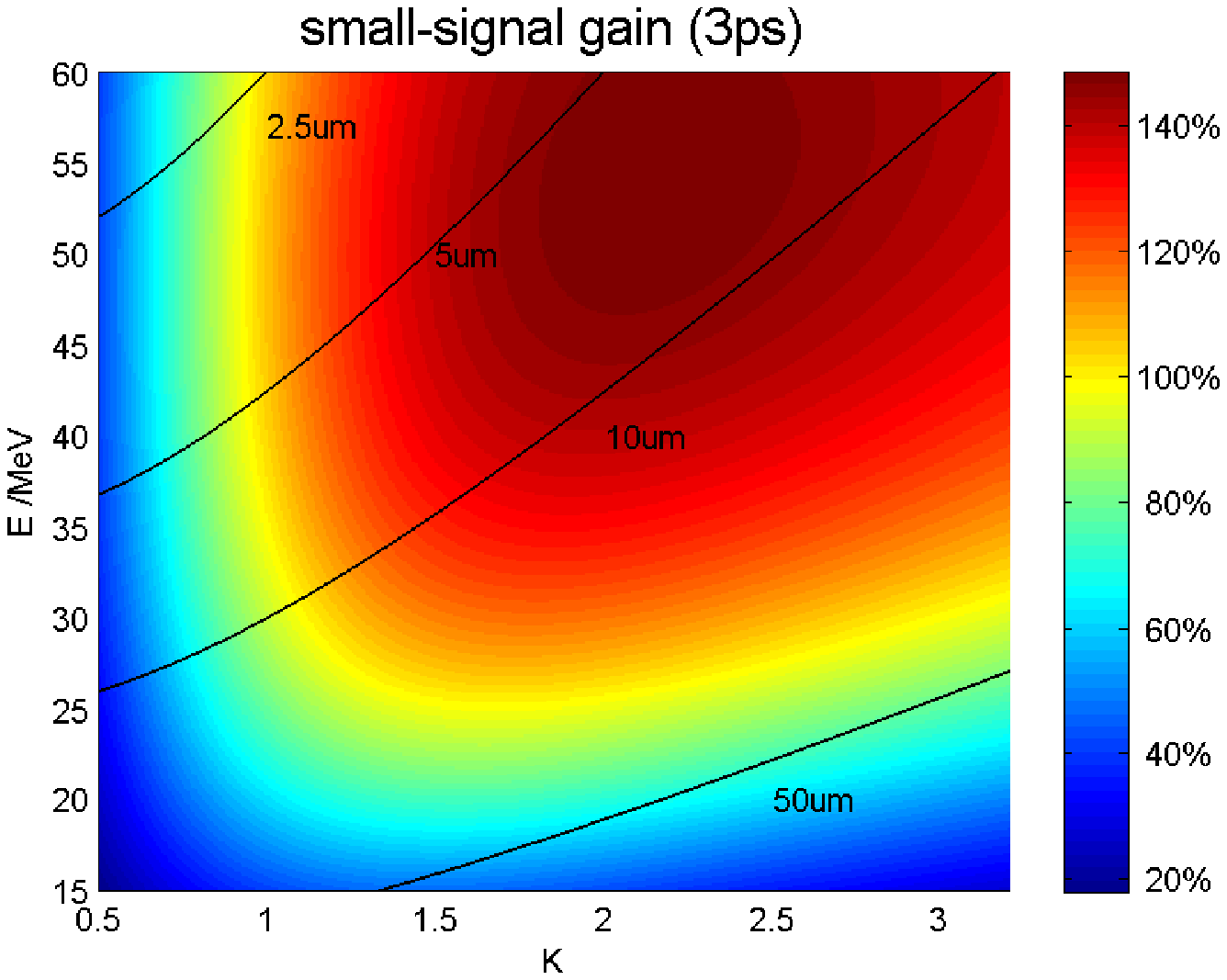}
\includegraphics[width=7cm]{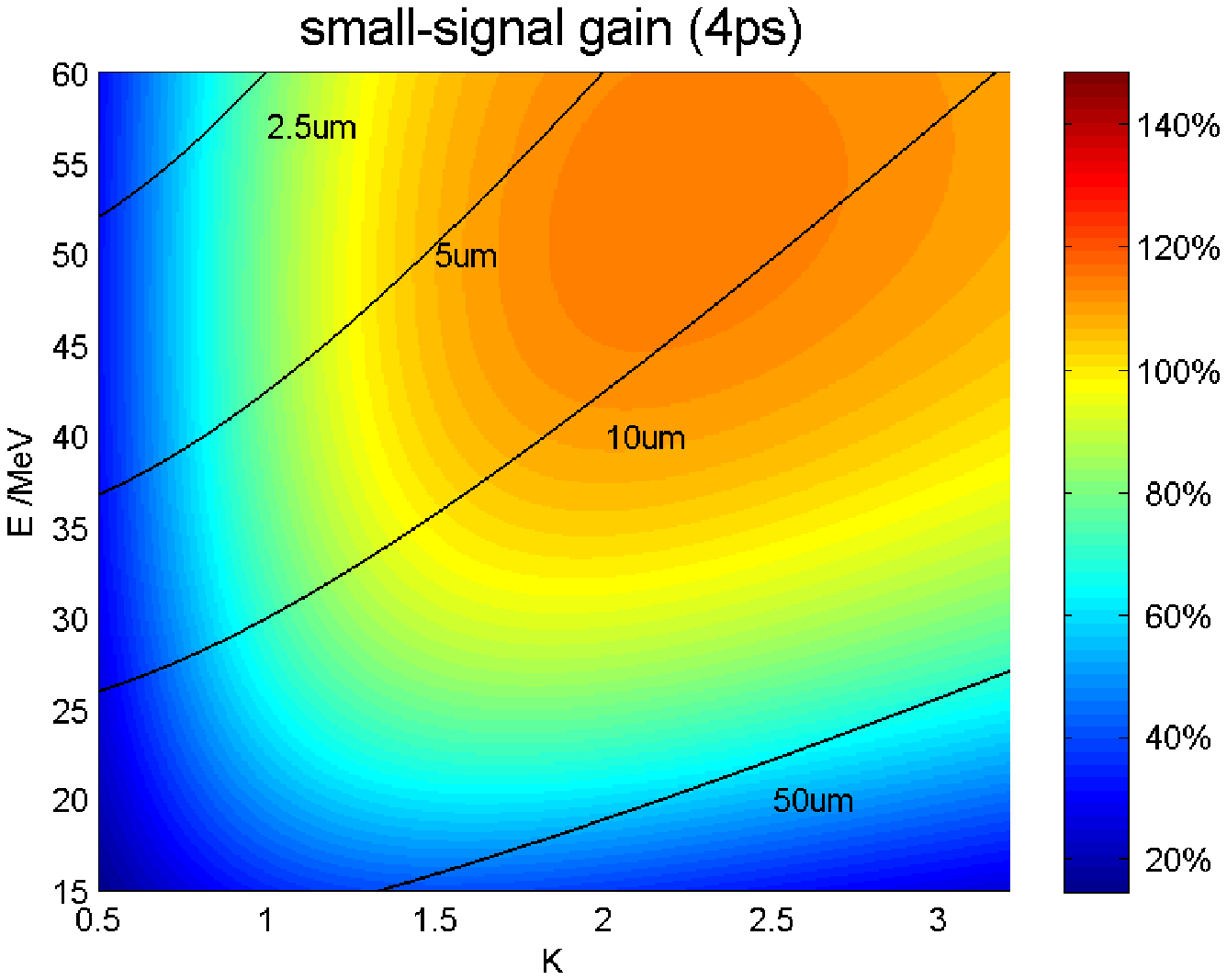}
\figcaption{\label{fig5}   The small signal of MIR-FEL. The energy spread and emittance are fixed to be 240 keV and 30 mm$\cdot$mrad, respectively. {\bf Left}: RMS electron bunch length of 3 ps; {\bf Right}: RMS electron bunch length of 4 ps.}
\end{center}

\begin{center}
\includegraphics[width=7cm]{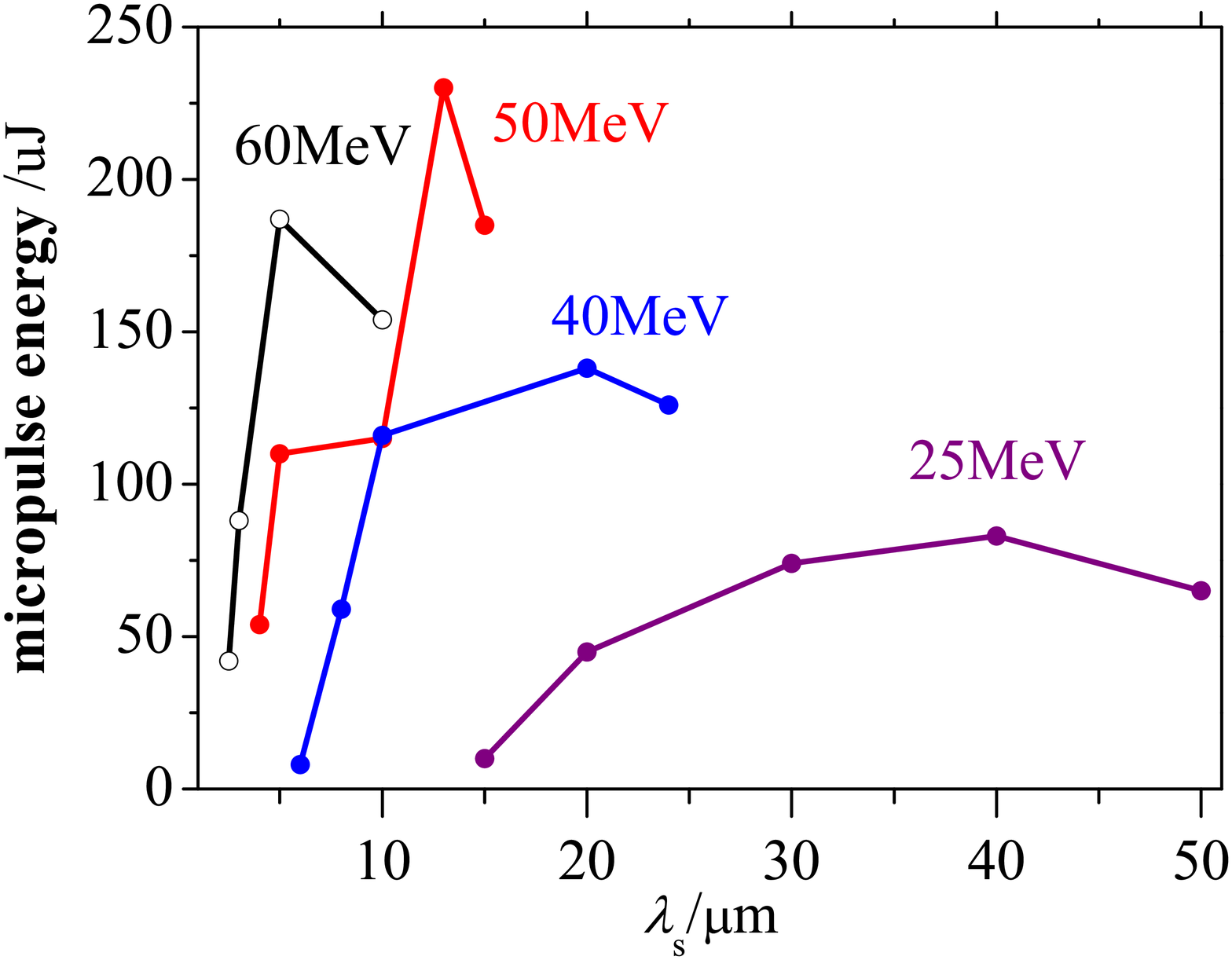}
\includegraphics[width=7cm]{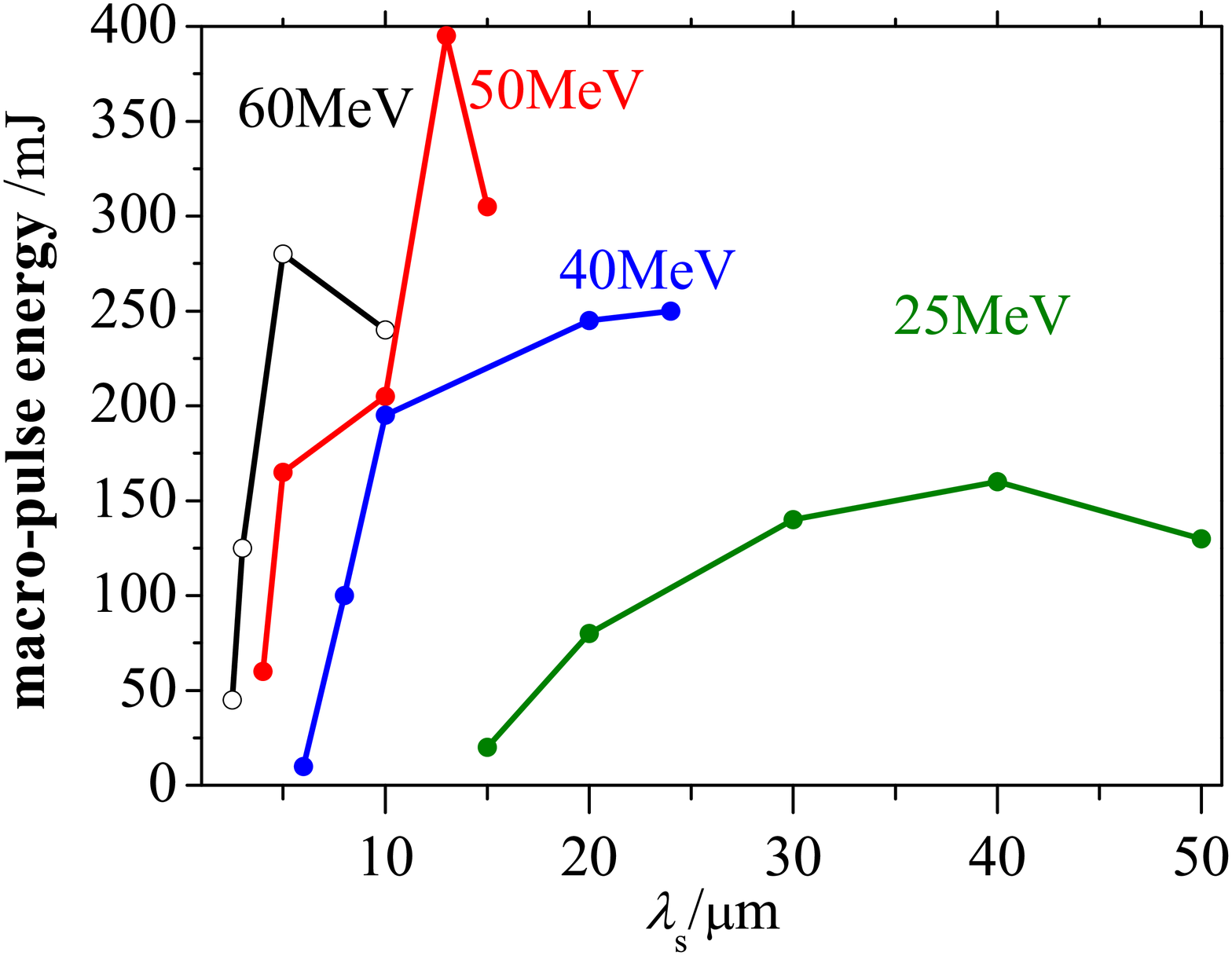}
\figcaption{\label{fig5}   Output pulse energy of MIR-FEL based on the electron microbunch repetition rate of 238 MHz and macrobunch length of 10 $\rm \mu$s. The energy spread and emittance are fixed to be 240 keV and 30 mm$\cdot$mrad, respectively. The RMS bunch length is 4 ps. {\bf Left}: micropulse; {\bf Right}: macropulse.}
\end{center}
\begin{multicols}{2}

 \subsection{MIR-FEL performance}
Based on the designed parameters in the previous Parts, the small signal gains of MIR-FEL are calculated and shown in Fig. 6. We can see that the gain is very high in the wavelength range of 4-30 $\rm{\mu}$m. For short wavelength (close to 2.5$\rm{\mu}$m), small $K$ leads to the low gain while for long wavelength (close to 50$\rm{\mu}$m), large relative energy spread and slippage effect are the reasons. In fact, from our Linac design, the energy spread at low energy is much smaller than 240 keV. That is to say, the gain in long wavelength should be much higher.

Figure 7 shows the output micropulse and macropulse energy of MIR-FEL simulated by OPC code. Note that the cavity length detuning is fixed to be two times radiation wavelength. One can find that the macro-pulse energy can exceed 100 mJ for most of the object wavelengths. Here, we simply choose four electron energies in the simulation. For specified spectral range of user requirements, we will find the optimum electron energy firstly, and the FEL performance will be better. We still can easily control the macro-pulse energy by selecting the electron microbunch repetition rate.

In addition, with the precise detuning of the cavity length , one can adjust the bandwidth of the output radiation. From the simulations, the radiation bandwidth can be tuned in 0.3-3\%.

\section{FIR oscillator}
The FIR oscillator is driven by the same Linac as MIR-FEL so that the electron beam quality is almost the same but a little worse. However, as the required electron beam energy is lower (15-25 MeV), the energy spread can be reduced to smaller than 150 keV. We may need to decompress the microbunch to suppress the slippage effect for long wavelength.

With the same considerations as MIR oscillator, the FIR undulator is optimized with period of 56 mm and total length of 2.24 m, and the cavity length is 5.04 m equal to MIR cavity length. The main difference from the MIR oscillator is that, due to the serious diffraction effect, we need to add a waveguide to reduce the diffraction loss in the FIR oscillator. A planar waveguide with the height of $b$=10 mm is used inside the undulator chamber. We calculate the small signal gain for the FIR-FEL as shown in Fig.8. The left figure is calculated by Eq.(3) for an open oscillator, while the right one is obtained by replacing the Gaussian mode of the optical beam by the fundamental waveguide mode. From Fig. 8, the waveguide enhanced the small signal gain by more than two times. In addition, according to the operation experience of CLIO and FELIX \cite{lab22,lab23}, "spectral gap" may appear in this waveguide FEL, and we are now considering this problem for FELiChEM.

\end{multicols}

\begin{center}
\includegraphics[width=7cm]{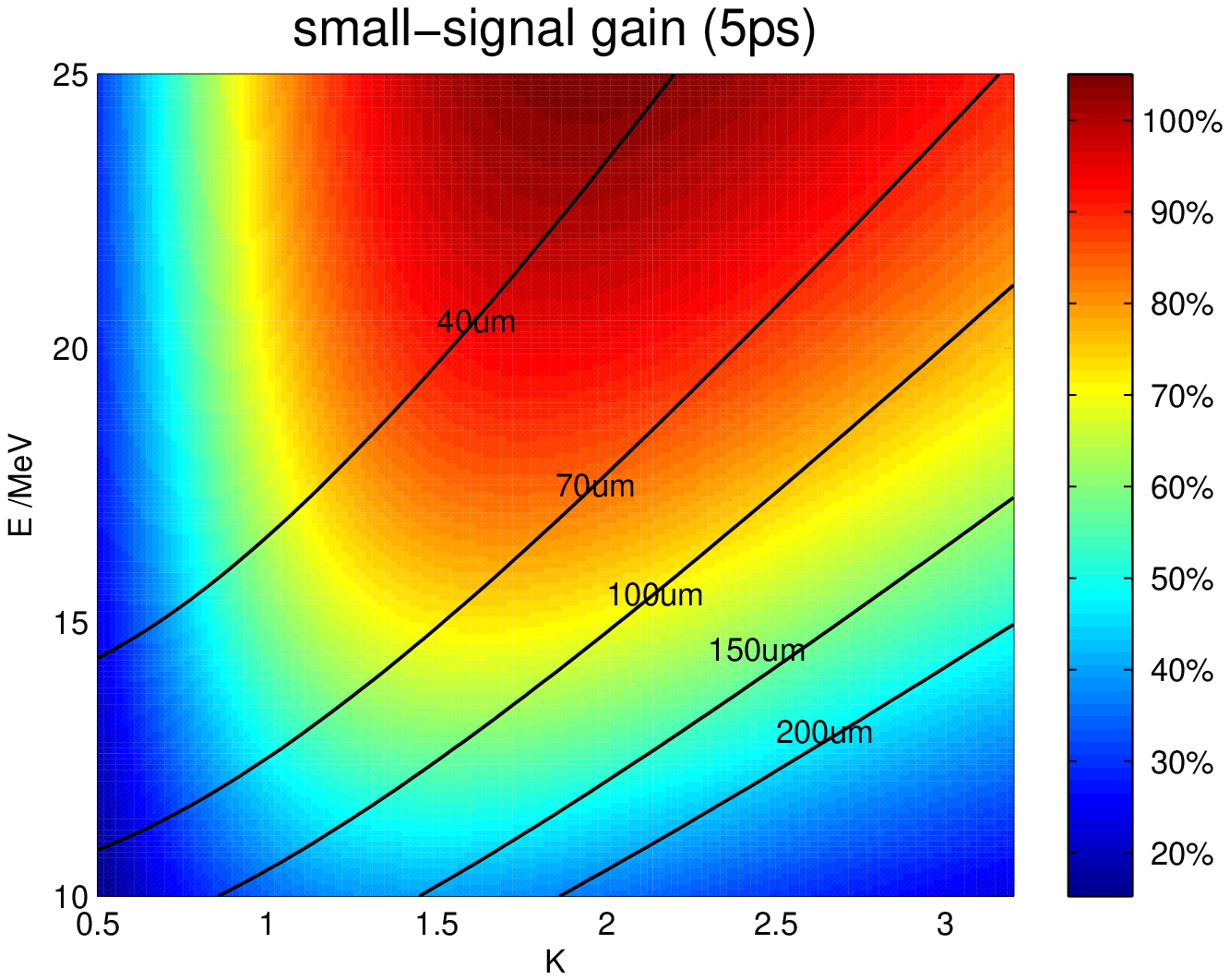}
\includegraphics[width=7cm]{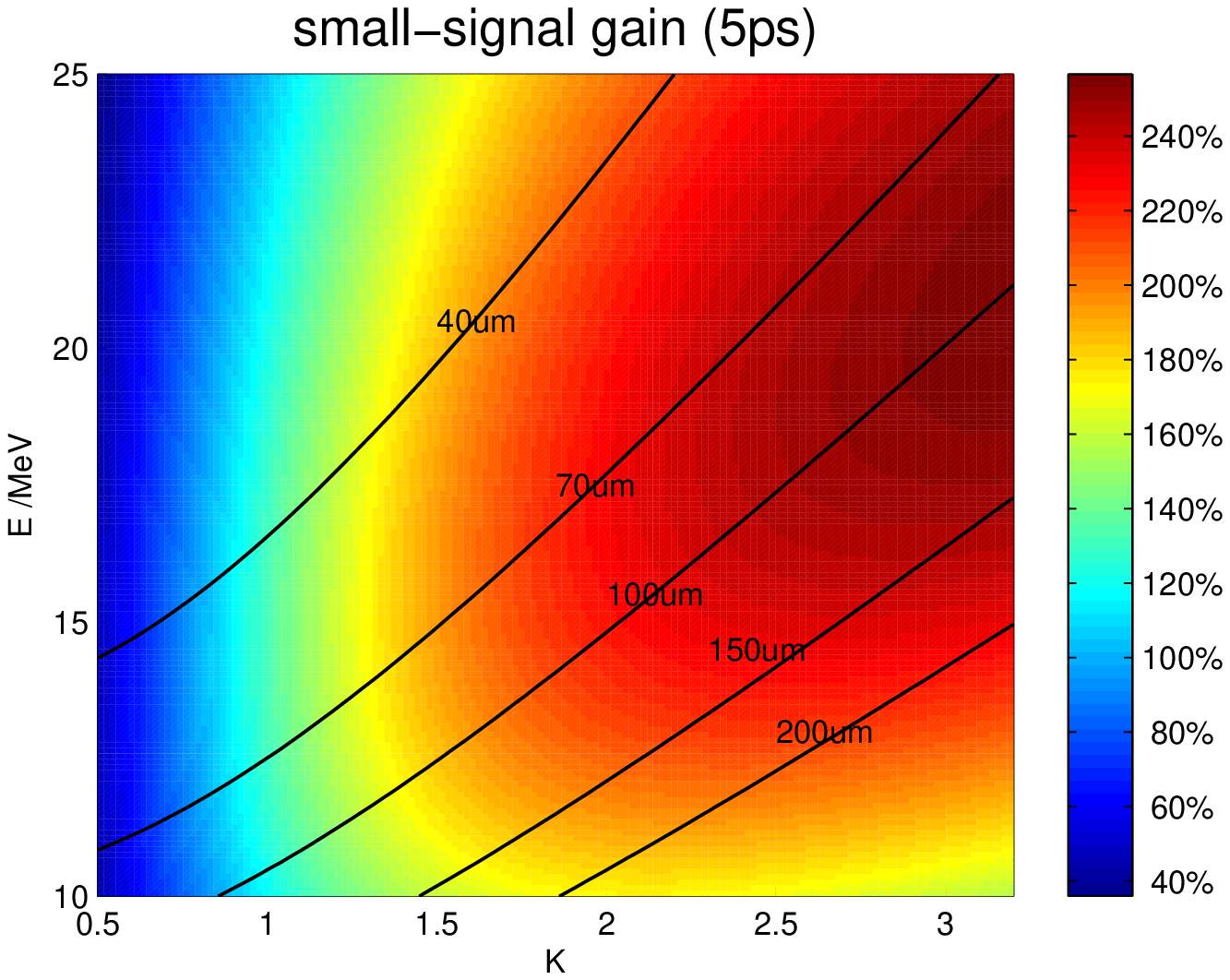}
\figcaption{\label{fig5}   The small signal of FIR-FEL. The energy spread and emittance are fixed to be 150 keV and 30 mm$\cdot$mrad, respectively. The electron bunch length is 5 ps. {\bf Left}: none waveguide; {\bf Right}: with a planar waveguide ($b$=10 mm).}
\end{center}
\begin{multicols}{2}

\begin{center}
\tabcaption{ \label{tab1}  Radiation performance of FELiChEM.}
\footnotesize
\begin{tabular*}{80mm}{c@{\extracolsep{\fill}}ccc}
\toprule Parameter  & Specification & Unit \\
\hline
Spectral range\hphantom{00}  & 2.5-200\hphantom{0} & $\rm{\mu}$m \\
Macropulse energy\hphantom{00} & 10-200\hphantom{0}& mJ   \\
Macropulse length\hphantom{0} & 5-10\hphantom{0} & $\rm{\mu}$s  \\
Macropulse repetition rate& 20 \hphantom{0} & Hz \\
Micro-pulse energy & 1-100\hphantom{0}& $\rm{\mu}$J  \\
Micro-pulse length & 1-5\hphantom{0}& ps  \\
Micro-pulse repetition rate & 476 /238 /119 /59.9 \hphantom{0}& MHz  \\
Bandwidth & 0.3\%-3\%\hphantom{0}&  - \\
Continuous tunability & 300\%-500\% \hphantom{0}& -  \\
\bottomrule
\end{tabular*}
\end{center}

\section{Summary}

In summary, we have introduced the FELiChEM that is building up an IR user facility in China. Brief physical design considerations and results are given and most of the basic and important parameters are determined. We summarize the anticipated radiation performance in Table 4, which make us believe that FELiChEM will build a high-level IR-FEL. In fact, the technical design of FELiChEM has been completed, and now most of the equipments are in fabrication or being procured. First light is targeted for the end of 2017.

\acknowledgments{The authors would like to thank the FHI, CLIO and FELIX people for generously sharing their valuable experiences in IR-FEL oscillator.}

\end{multicols}

\vspace{-1mm}
\centerline{\rule{80mm}{0.1pt}}
\vspace{2mm}

\begin{multicols}{2}

\end{multicols}

\clearpage
\end{CJK*}

\begin{thebibliography}{90}

\vspace{3mm}

\bibitem{lab1} S. Krishnagopal, V. Kumar, Radiation Physics and Chemistry {\bf 70}, 559---569 (2004).

\bibitem{lab2} J.M. Ortega, M. Bergher, R. Chaput et al, Nucl. Instrum. Methods A {\bf 285}, 97---103 (1989).

\bibitem{lab3} J.M. Ortega, Nucl. Instrum. Methods A {\bf 341}, 138---141 (1994).

\bibitem{lab4} W. Schollkopf et al, First lasing of the IR FEL at the Fritz-Haber Institut, Berlin. Proc. of FEL Conference. Nara, Japan. 2012, p. 1---4.

\bibitem{lab5} W. Schollkopf et al, The new IR and THz FEL Facility at the Fritz Haber Institute in Berlin. Proc. of SPIE {\bf 9512}, 95121L-1 (2015).

\bibitem{lab6} P.W. van Amersfoort, R.W.B. Best, R. van Buuren et al, Nucl. Instrum. Methods A {\bf 296}, 217---221 (1990).

\bibitem{lab7} P.W. van Amersfoort, R.J. Bakker, J.B. Bekkers et al, Nucl. Instrum. Methods A {\bf 318}, 42---46 (1992).

\bibitem{lab8} Hai-Xiao Deng et al, Chinese Physics C (HEP \& NP), {\bf 38}, 028101 (2014).

\bibitem{lab9} Heting Li et al, J. of IRMM\&THz Waves, 2015, DOI: 10.1007/s10762-016-0258-9, to be published.

\bibitem{lab10} N. Vinokurov, J. of IRMM\&THz Waves, {\bf 32}, 1123-1143 (2011).

\bibitem{lab11} Y. C. Huang, Appl. Phys. Lett. \textbf{96}, 2315039 (2010).

\bibitem{lab12} Jia-Lin Xie et al, Chinese Physics C (HEP \& NP), {\bf 18}, 572-576 (1994) (in Chinese).

\bibitem{lab13} Yuan-Zhang Wang, Xiao-Jian Shu. High Power Laser and Particle Beams, {\bf 7}, 501-506 (1997) (in chinese).

\bibitem{lab14} Yu-Huan Dou et al, High Power Laser and Particle Beams, {\bf 25}, 662-666 (2013) (in chinese).

\bibitem{lab15} Feng Luo, Hua Bei, Zhi-Min Dai, Chinese Physics C (HEP \& NP), {\bf 34}, 512-515 (2010).

\bibitem{lab16} DAI Jin-Hua, DENG Hai-Xiao, DAI Zhi-Min, Chinese Physics C, {\bf 36}, 648-652 (2012).

\bibitem{lab17} G.Dattoli et al, IEEE J. Quantum Electron, {\bf 17}, 1371 (1981).

\bibitem{lab18} G.Dattoli et al, IEEE J. Quantum Electron, {\bf 20}, 637 (1984).

\bibitem{lab19} P. J. M. van der Slot, H. P. Freund, W. H. Miner Jr., S.V. Benson, M. Shinn, K.-J. Boller, Phys. Rev. Lett. {\bf 102}, 244802 (2009).

\bibitem{lab20} S. Reiche, Nucl. Instrum. Methods A, {\bf 429}, 243 (1999).

\bibitem{lab21} W. Schollkopf, S. Gewinner, W. Erlebach et al, The IR and THz FEL at the Fritz-Haber Institut. Proc. of FEL Conference. New York, USA. 2013. p. 657---660.

\bibitem{lab22} Ruslan Chulkov, Vitaliy Goryashko, Denis D. Arslanov et al, Phys. Rev. ST Accel. Beams {\bf 17}, 050703 (2014).

\bibitem{lab23} J.-M. Ortega, J.-P. Berthet, F. Glotin, and R. Prazeres, Phys. Rev. ST Accel. Beams {\bf 17}, 100701 (2014).

\end{thebibliography}
\end{document}